# Doping-Induced Electronic and Structural Phase Transition in the Bulk Weyl Semimetal Mo$_{1-x}$W$_x$Te$_2$

O. Fedchenko[1], F. K. Diekmann[2], P. Rüßmann[3,4], M. Kallmayer[5], L. Odenbreit[1], S. M. Souliou[6], M. Frachet[6], A. Winkelmann[7], M. Merz[6], S. V. Chernov[8,9], D. Vasilyev[1], D. Kutnyakhov[10], O. Tkach[1], Y. Lytvynenko[1,11], K. Medjanik[1], C. Schlueter[8], A. Gloskovskii[8], T. R. F. Peixoto[8], M. Hoesch[8], M. Le Tacon[6], Y. Mokrousov[1,4], K. Rossnagel[2,12], G. Schönhense[1], H.-J. Elmers[1]

[1] Institut für Physik, Johannes Gutenberg-Universität Mainz, 55128 Mainz, Germany
[2] Institut für Experimentelle und Angewandte Physik, Christian-Albrechts-Universität zu Kiel, 24098 Kiel, Germany
[3] Institute for Theoretical Physics and Astrophysics, University of Würzburg, 97074 Würzburg, Germany
[4] Peter Grünberg Institut and Institute for Advanced Simulation, Forschungszentrum Jülich and JARA, 52425 Jülich, Germany
[5] Surface Concept GmbH, Am Sägewerk 23a, 55124 Mainz, Germany
[6] Institute for Quantum Materials and Technologies, Karlsruhe Institute of Technology, 76021 Karlsruhe, Germany
[7] Academic Centre for Materials and Nanotechnology, AGH University of Science and Technology, Kraków, Poland
[8] Deutsches Elektronen-Synchrotron DESY, 22607 Hamburg, Germany
[9] Department of Physics and Astronomy, Stony Brook University, Stony Brook, USA
[10] Deutsches Elektronen-Synchrotron DESY, Center for Free-Electron Laser Science, 22607 Hamburg, Germany
[11] Institute of Magnetism of the NAS of Ukraine and MES of Ukraine, 03142 Kyiv, Ukraine
[12] Ruprecht Haensel Laboratory, Deutsches Elektronen-Synchrotron DESY, 22607 Hamburg, Germany

A comprehensive study of the electronic and structural phase transition from $1T$ to $T_d$ in the bulk Weyl semimetal Mo$_{1-x}$W$_x$Te$_2$ at different doping concentrations has been carried out using time-of-flight momentum microscopy (including circular and linear dichroism), X-ray photoelectron spectroscopy (XPS), X-ray photoelectron diffraction (XPD), X-ray diffraction (XRD), angle-resolved Raman spectroscopy, transport measurements, density functional theory (DFT) and Kikuchi pattern calculations. High-resolution angle-resolved photoemission spectroscopy (ARPES) at 20 K reveals surface electronic states, which are indicative of topological Fermi arcs. Their dispersion agrees with the position of Weyl points predicted by DFT calculations based on the experimental crystal structure of our samples determined by XRD. Raman spectroscopy confirms the inversion symmetry breaking for the $T_d$-phase, which is a necessary condition for the emergence of topological states. Transport measurements show that increasing the doping concentration from 2 to 9% leads to an increase in the temperature of the phase transition from $1T$ to $T_d$ from 230 K to 270 K. Magnetoresistance and longitudinal elastoresistance show significantly increased values in the $T_d$-phase due to stimulated inter-pocket electron backscattering. The results demonstrate the close relationship between electronic properties and elastic deformations in MoTe$_2$.

## I. INTRODUCTION

Transition metal dichalcogenides (TMDCs) have attracted the attention of the research community due to their highly interesting electronic properties, structural chemistry, topological electronic states [1-5] and wide range of potential applications [6]. Large magnetoresistance, topological Weyl states and pressure-induced superconductivity have been reported [7-8]. Partial substitution of elements in TMDCs [9] and thickness engineering with atomic precision [10] provide tuning knobs for the variation of physical properties.

MoTe$_2$ is unique among the TMDCs because it can be grown in both semiconducting 2$H$ (space group P6$_3$/mmc) and semi-metallic 1$T'$ (space group P2$_1$/m) forms. Thus, MoTe$_2$ offers the possibility to control the transitions between these phases with temperature, electrical or chemical doping, laser irradiation, tensile strain, electric field, terahertz light field, electron beam irradiation, or plasma treatment [11-16]. In addition, *ab-initio* calculations of the effect of transition metal doping in MoTe$_2$ predict spontaneous magnetic order [17,18].

2$H$-MoTe$_2$ is a semiconductor in which the bands are spin-degenerate due to the spatial inversion symmetry of the crystal structure [19]. Zhang *et al.* [19] reported a resistive memory device based on an electric field-induced reversible structural transition for vertical 2$H$-MoTe$_2$ [14]. The realisation of the ohmic contact between heterophase-homojunctions for in-plane 2$H$/1$T'$-MoTe$_2$ heterostructures was reported in Refs. [20, 21]. Mleczko *et al.* demonstrated the unipolar n-type operation in 2$H$-MoTe$_2$-transistors [22]. Therefore, 2$H$-MoTe$_2$ is a promising candidate for applications in lasers, logic circuits, photodetectors, and sensors. A possible practical implementation of MoTe$_2$ is an atomically thin phase-change memory device, which offers favourable switching properties due to the small energy difference between 2$H$ and 1$T'$ phases (about 35 meV) [23], [24].

The 1$T'$-MoTe$_2$ is a metastable phase that can be driven by cooling into the orthorhombic $T_d$ phase (space group Pmn2$_1$) with broken inversion symmetry, as demonstrated by structural, electrical, and optical measurements [25]. In the $T_d$ – phase, MoTe$_2$ is a type-II topological Weyl semimetal with potential applications in spintronics and valleytronics [26]. The highly symmetric semiconducting phase of MoTe$_2$ (centrosymmetric hexagonal 2H) shows bands with high spin polarisation up to 80% at the surface ($K$ versus $K'$ valleys) along $k_Z$ (existence of the intrinsic hidden spin polarisation in centrosymmetric materials), while in-plane measurements at the $K$ and $K'$ points show negligible spin polarisation [27]. The non-centrosymmetric lattice structure is a necessary condition for the existence of topological Weyl states (WS). Li *et al.* reported spin- and momentum-resolved photoemission measurements for $T_d$ – MoTe$_2$ [23], where circular dichroism was used to obtain a direct fingerprint of the chiral quasiparticle states. Spin-resolved measurements of the Fermi surface revealed a spin texture of Weyl cones related to the chiral charge of the Weyl points (WPs). The Weyl phase is associated with the WPs in MoTe$_2$, but this issue is still under discussion [28,29]. The number of WPs depends on the lattice constant, the strength of the spin-orbit coupling and the positions of the atoms [30,31].

In contrast to MoTe$_2$, WTe$_2$ exhibits a stable orthorhombic $T_d$ -phase independent of temperature, with topological Weyl-states and Fermi arcs that are smaller than in case of MoTe$_2$ [32, 33]. The orthorhombic phase in WTe$_2$ is very stable against external stimuli.

Here, we explore the possibility of combining the ability to stimulate reversible phase changes between topological and trivial phases with the large energy splitting of the WPs [9], [34] by a partial replacement of Mo ions with W ions in MoTe$_2$. By increasing the doping x in Mo$_{1-x}$W$_x$Te$_2$, the crystal structure switches from the 2$H$-phase, via the 1$T'$- to the $T_d$ -phase [35]. The resulting phase diagram is ideal for exploring the functional response of the electronic properties and verifying the presence of the Weyl phase [31]. By varying the W concentration, the distance between Weyl points in momentum space can be tuned [30]. In addition, the ground state energy difference between semiconducting and semimetallic phases can be tuned [24]. While pristine 1$T$`-MoTe$_2$ has many bands crossing the Fermi level along the Γ-X and Γ-Z directions, the Fermi level of Mo$_{1-x}$W$_x$Te$_2$ shows only few band crossings similar to WTe$_2$. Depending on the doping (W) concentration, the *k*- and energy-space distance between a pair of Weyl nodes and the Fermi arc lengths vary [30]. It is predicted that if the doping concentration exceeds the critical value, the touching point splits into a pair of Weyl nodes of opposite chirality [30] connected by a Fermi arc. Thus, the length of the Fermi arc connecting two Weyl nodes is a measure of the topological strength of the system. Doping can be used to tune the topological strength of the system, revealing its significant advantage over other TMDCs.

In this article, we demonstrate the occurrence of a reversible structural phase transition between two stable phases, the $T$' and the $T_d$ -phase, in W-doped MoTe$_2$ (see Fig.1). Using momentum microscopy, we show that the phase transition from trivial to topological electronic states does indeed occur. In addition, we have performed a comprehensive set of experimental studies including Raman spectroscopy, X-ray diffraction (XRD), X-ray photoelectron diffraction (XPD), and electronic transport to characterise the structural changes in the samples. Using high-resolution photoemission spectroscopy in combination with ab-initio calculations, we find evidence for the appearance of Fermi arcs connecting pairs of Weyl-nodes in the occupied and unoccupied parts of the electronic system in the $T_d$ -phase.

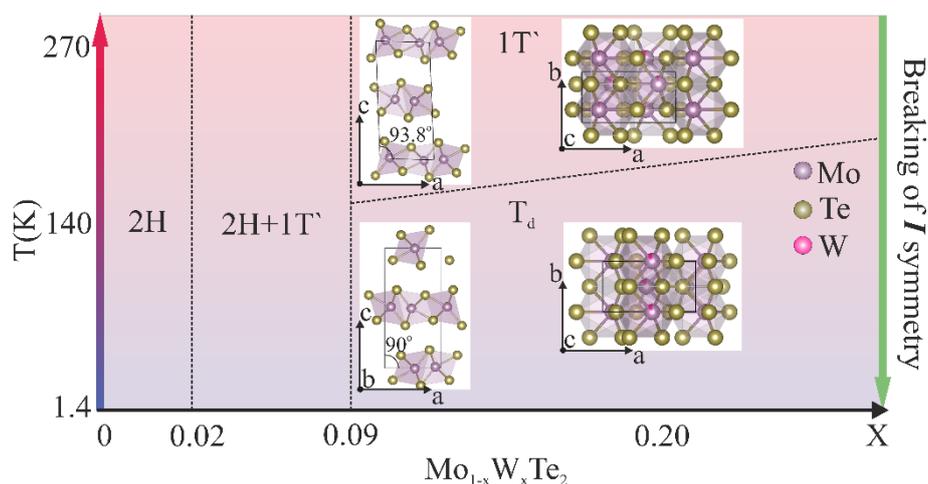

**Figure 1.** Phase diagram for Mo$_{1-x}$W$_x$Te$_2$ based on Raman spectroscopy and XRD measurements. Precise atomic positions are experimentally determined.

The article is organized as follows: After the introduction of experimental and theoretical methods, we discuss the valence band photoemission results as the main topic. Then, we present a detailed structural analysis based on Raman spectroscopy, X-ray diffraction, and X-ray photoelectron diffraction. Finally, we discuss electronic transport properties as a function of temperature, magnetic field, and strain.

## II.  Experimental and theoretical methods

Single crystals of Mo$_{1-x}$W$_x$Te$_2$ were grown by the standard chemical vapor transport method: A near-stoichiometric mixture of high-purity Mo, W and Te was placed in a quartz ampoule together with iodine (5 mg/cm$^3$) as transport agent; the ampoule was sealed and heated in a four-zone furnace under a temperature gradient of 890-710 °C within 336 h. The formation of polytypes was prevented by quenching of the hot ampoules in water. Typical crystal sizes were about 0.3×0.3×0.1 mm$^3$. The quality of the samples was confirmed by XRD and Raman spectroscopy (see Supplementary Material, Fig. S1).

The photoemission experiments were performed with a time-of-flight momentum microscope (ToF-MM) [36], and a single-hemisphere momentum microscope [37]. Photoelectrons were excited by a He discharge lamp (21.2 eV), by a pulsed laser (6.4 eV, 80 MHz repetition rate, APE), and by soft X-ray and hard X-ray radiation (beamlines P04 and P22 correspondingly, PETRA III, DESY, Hamburg). Photoemission experiments at 21.2 eV were performed using the high-resolution single-hemisphere momentum microscope with the energy resolution set to 10 meV. For the soft X-ray experiments, we used a time-of-flight momentum microscope at the open port I of beamline P04. The total energy resolution was 60 meV.

Using X-ray excitation, direct transitions in three-dimensional momentum space result in the photoelectron intensity distribution I($E_B$, $k_x$, $k_y$, $k_z$), which is simultaneously recorded in the time-of-flight momentum microscope (ToF MM). I($E_B$, $k_x$, $k_y$, $k_z$) represents the spectral density function modulated by matrix elements accounting for the photo-excitation probability for a given initial $k_i$ and final $k_f$.

The photoemission experiments were performed with the incidence angle of the photon beam at 22° with respect to the sample surface. Azimuthal orientation of the sample was adjusted such that the photon incidence plane coincided with the *c*-axis [001]. Prior to the photoemission experiments, the samples were cleaved *in situ* in ultrahigh vacuum by the tape method. The base pressure in the microscope chamber at room temperature was 3·10$^{-10}$ mbar.

Angle-resolved Raman and X-ray diffraction (XRD) investigations were performed at KIT (Karlsruhe) in the temperature range from RT down to 1.4 K. Raman scattering experiments were performed on crystals of Mo$_{1-x}$W$_x$Te$_2$ using a single grating Jobin-Yvon LabRam HR evolution spectrometer in backscattering geometry. The 633 nm line of a He-Ne laser was used as excitation. Low temperature conditions were realized using a continuous flow liquid helium cryostat. The experimentally determined structural parameters from XRD measurements were used for density functional theory (DFT) band structure and orbital-projected density-of-states (DOS) calculations, as well as X-ray photoelectron diffraction (XPD) pattern calculations based on the Bloch-wave approach.

Angle-resolved polarized Raman spectroscopy can be utilized to assign the Raman modes based on crystal symmetry and Raman selection rules and also to characterize the crystallographic orientation of anisotropic materials [38], for example low-symmetry 2D materials [39, 40].

The DFT simulations in this work were performed using the full-potential relativistic Korringa-Kohn-Rostoker Green's Function (KKR) method [41] as implemented in the JuKKR code package [42]. We used the Local Density Approximation (LDA) as the parameterisation for the exchange correlation functional [43] and applied an $l_{max} = 3$ cutoff in the angular momentum expansion of the Green's function with an exact partitioning of the unit cell into atomic cells [44, 45].

The calculations for the 2*H* crystal structures of Mo$_x$W$_{1-x}$Te$_2$ were performed starting from the bulk crystal structures of MoTe$_2$ [46] and WTe$_2$ [47] from the Materials Project database [48]. Alloys of 2*H*-Mo$_{1-x}$W$_x$Te$_2$ were computed within the Coherent Potential Approximation (CPA) [49], where the crystal structure was interpolated between the clean 2*H* phases of MoTe$_2$ and WTe$_2$. The CPA calculations for the $T_d$ and 1*T'* structures of Mo$_{0.88}$W$_{0.12}$Te$_2$ started from the crystal structure determined by XRD at 100 K and 300 K, respectively.

Electrical resistance measurements were performed in a four-probe geometry. Magnetic field was applied along the crystallographic *c*-axis. Strain was applied by gluing the Mo$_{1.91}$W$_{0.09}$Te$_2$ sample to a piezoelectric stack (Pst 150/5*5*7 from Piezomechanik) using DevCon 5min epoxy [50].

## III. Photoemission Results
### a. MoTe$_2$ and Mo$_{0.98}$W$_{0.02}$Te$_2$

S.M. Oliver and colleagues [34] reported that Mo$_{1-x}$W$_x$Te$_2$ exhibits a stable 1$T$-phase with a W content x ≤ 0.4 based on XRD and STEM measurements [51], and a $T_d$ - phase for $x ≥ 0.63$. A two-phase mixed phase (1$T$ and $T_d$) exists for the 0.04 < $x$ < 0.63 range. It has also been found that annealing in vacuum at 750° C for 72 hours followed by slow cooling back to room temperature can induce a reversible phase transition from the metallic 1$T$ phase to the semiconducting 2$H$ phases at $x ≈ 0.09$ [9,34,35].

The phase diagram, based on the Raman spectroscopy and XRD measurements, differs slightly from previous reports [34,35] (see Fig.1), where the authors claimed that pristine MoTe$_2$ can be obtained in both 2$H$ and 1$T$ phases using different growth procedures. In our case, even after rapid quenching, Mo$_{1-x}$W$_x$Te$_2$ samples with $x$=0 and 0.02 show a 2$H$ phase (hexagonal) which is stable and does not show any phase transitions within the investigated temperature range. Samples with x=0.09 show 1$T$-phase (monoclinic) above 150 K and $T_d$ – orthorhombic phase with broken inversion symmetry at low temperatures (LT, 20K).

Photoemission spectroscopy using ToF MM in the laboratory and hard X-ray photoemission spectroscopy at a synchrotron confirm that pristine MoTe$_2$ is in the 2$H$ phase as expected and shows no phase transitions at low temperatures. Only a continuous thermal shift of the Fermi level can be observed. These results are in good agreement with previously published results [51] (stating that the 2$H$ phase is thermally stable), as well as with temperature-dependent Raman spectroscopy and XRD measurements.

X-ray photoemission spectroscopy measurements performed at the hard X-ray beamline P22 at PETRA (DESY, Hamburg) showed that the valence band maximum of 2$H$-MoTe$_2$ exhibits a spectral shift of 100 meV between LT and RT. The shift amounts to approximately 0.4 meV/K. This effect is reversible and reproducible (see Supplementary Material, Fig. S2). We attribute this spectral shift to defect bands. Defect bands associated with Te vacancies appear within the gap of 2$H$-MoTe$_2$, thereby reducing the band gap and shifting the Fermi level to the valence band edge [53]. The authors of Ref. [53] showed theoretically and experimentally that the Fermi level is shifted by +250 meV from LT to RT for the 2$H$ phase in the case of 6% Te vacancies.

Mo$_{0.98}$W$_{0.02}$Te$_2$ shows a similar behaviour as pristine MoTe$_2$. The three-fold symmetry of the band structure [see Figs. 2(a,b) and Fig. S4] is indicative for the 2$H$ phase. The 2$H$ phase also shows a semiconducting state, where the Fermi level is again defined by impurity states below the conduction band edge, resulting in a very low density of states at the Fermi level. According to the calculated DOS, the semiconducting energy gap is about 1 eV (see Fig. S3).

### b. Mo$_{0.91}$W$_{0.09}$Te$_2$

To obtain the stable monoclinic 1$T$'-phase at 300 K (RT), the W concentration had to be further increased. The 1$T$'-phase can be stabilized with a W concentration of 9%, as confirmed by Raman spectroscopy and XRD measurements at RT (see below). At this W concentration, the constant energy maps show a two-fold symmetry [Figs. 2(c,d)], in contrast to the three-fold symmetry observed for lower W concentration [Figs. 2(a.b)]. The 1$T$'-phase is metallic. In this case, the density of states is significantly larger at the Fermi level [Fig. 3(c)].

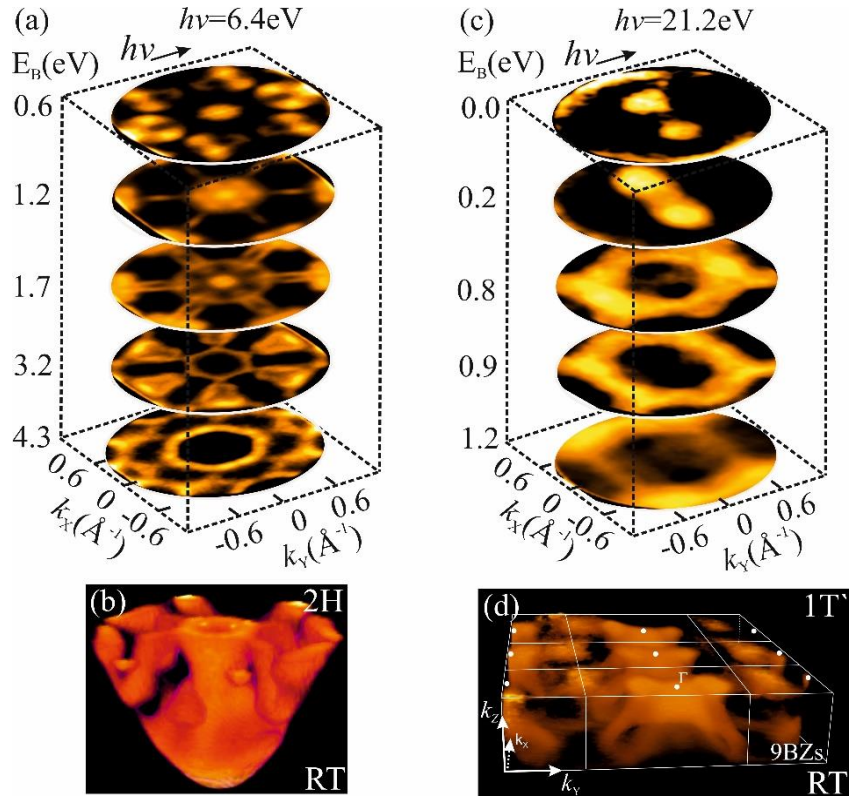

**Figure 2.** (a) $k_x$ - $k_y$ momentum distributions at the indicated binding energies $E_B$, obtained from the 3D data array of $Mo_{0.98}W_{0.02}Te_2$. (b) Three-dimensional representation of the photoemission intensity $I(E_B,k_x,k_y)$ for $Mo_{0.98}W_{0.02}Te_2$. (c,d) Similar data for $Mo_{0.91}W_{0.09}Te_2$. All data was measured at RT. Arrows in (a,c) denote the photon-impact direction.

According to XRD, Raman and photoemission spectroscopy measurements, the compound $Mo_{1-x}W_xTe_2$ with x=0.09 exhibits a phase transition from $1T'$ to $T_d$ around 170 K. Precise measurements of the VB and core levels (Mo 3$d$, Te 3$d$, Te 4$p$) were performed during heating and cooling cycles between RT and 20 K. In the case of high-energy XPS spectra (recorded at hv = 3400 eV), the Shirley background was subtracted, and asymmetric spectra were fitted with the Lorentz distribution and the Doniach-Sunjic profile. Both fit functions showed identical peak positions. It was found that the VB region does not show any temperature-induced shifts. This can be expected because the system is metallic. In contrast, the Mo 3$d$ core level shows a shift at the phase transition [see Fig. 3(b)]. Interestingly, this shift of the peak position is significantly larger for the Mo 3$d$ core level than for the Te core levels (see Supplementary Material Fig. S2). The observed core-level shift hints to a change of the screening effect of the valence band and thus to a change of the electronic structure at the phase transition. The electronic states responsible for the screening effect are localized mainly in the planes comprising the Mo atoms.

The electronic structure of $1T'$-$Mo_{1-x}W_xTe_2$ is considerably different from that of $2H$-$MoTe_2$. The finite DOS at $E_F$ indicates a metallic phase and is consistent with transport measurements (metallic behaviour of resistivity) [see Fig. 3(c)]. The DOS around $E_F$ is mainly derived from Mo 4$d$ (and correspondingly W 5$d$) and Te 5$p$ orbitals (see Supplementary Material Fig. S3). Mo electron-like and Te hole-like bands cross each other near $E_F$, resulting in a band inversion. The orbital projected DOS (see Supplementary Material, Fig. S3) shows that the main contribution at $E_F$ originates from Mo (W) $d$ orbitals for all three phases: $2H$, $1T'$ and $T_d$. At high binding energies (>10 eV), the opposite is true, with Te $p$ states largely dominating.

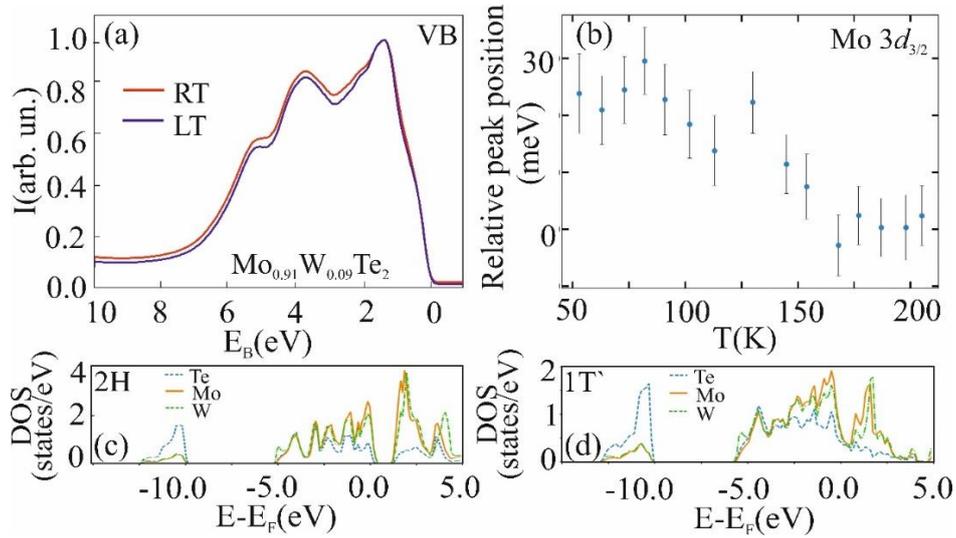

**Figure 3.** (a) Momentum-integrated photoelectron spectra for the valence band measured using hard X-rays (hv = 3400 eV) at RT and LT for $Mo_{0.91}W_{0.09}Te_2$. (b) Shift of the Mo $3d_{3/2}$ core-level peak with temperature for $Mo_{0.98}W_{0.02}Te_2$. (c) Calculated density of states for $2H$ and $1T'$-$Mo_{1-x}W_xTe_2$ (x=0.02; 0.09).

As was mentioned above, the $1T'$ phase changes to the orthorhombic $T_d$ phase at low temperature (see Fig. 1). Figs. 4(a,b) show constant energy maps for the high- and low-temperature phases. These photoemission measurements were carried out with the single-hemisphere momentum microscope and with an energy resolution set to about 10 meV, and constant energy maps were acquired in steps of 5 meV to allow the detection of the surface states. The $1T'$ phase at RT shows two distinct regions of high photoemission intensity that are related to the band minima of two electron pockets. At low temperature, additional photoemission intensity appears between the two electron pockets. This can be observed in the constant energy maps [Figs. 4(e,f) and in the corresponding intensity profiles [Fig. 4(d)]. The additional intensity is attributed to the hole pocket that is shifted to smaller binding energy in the $T_d$ phase, and thus crosses the Fermi level. This effect is illustrated in the schematic energy-band representation of valence and conduction bands at RT and LT in Figs. 4(i,j).

The $E_B$ versus parallel momentum cuts through the data array [Figs. 4(g,h)] show distinct band dispersions for the $1T'$ and $T_d$ phases. The $T_d$ phase exhibits a pair of split-off bands, separating from the hole band at $E_B = 75$ meV (marked with a white arrow). This state does not appear in the $1T'$ phase.

We attribute this split-off band to a topological surface state (TSS) connecting the pair of Weyl nodes above and below the Fermi level. An experimental proof of this assumption would be spin-resolved measurements, which are beyond the scope of this work. However, further insight can be gained by the ab-initio calculations: In contrast to the $1T'$-phase with spatial inversion symmetry, the $T_d$ phase exhibits crystal inversion symmetry breaking. Consequently, the electronic structure shows a significant band splitting except for the time-reversal invariant points (see Supplementary Material Fig. S4(e)).

This interpretation is consistent with previous reports [54] showing that the type-II nature of the Weyl cones results in the coexistence of projected bulk and surface states. The bulk state consists of one hole pocket in the center of the BZ and two elliptical electron pockets near the left and right edges. The projected Weyl nodes are located at the boundary of the hole and electron pockets. Theoretical analysis has revealed that the Fermi-arc surface states are derived from the Mo-$d$ orbitals.

Ref. [54] reports that the surface states forming the Fermi arcs, which stem from the projected Weyl cone pocket and then vanish into the bulk continuum, do not contribute to the observable quantum interference patterns. Due to the tilted nature of the type-II Weyl cone, the area of the projected Weyl bulk states is much larger than for type-I Weyl cones, and therefore the topological sink effect is much more pronounced for type-II Weyl states.

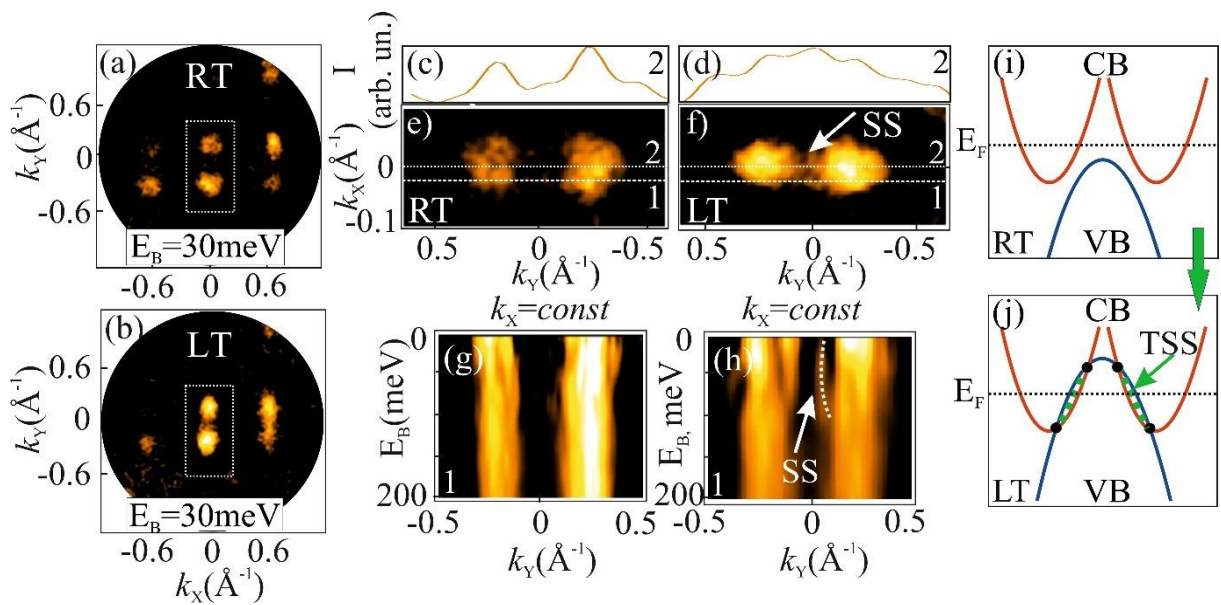

**Figure 4.** Constant-energy momentum distributions (a,b,e,f), intensity profiles (c,d) and $E_B$-versus-$k_\parallel$ (g,h) sections through the 3D data array of $Mo_{0.91}W_{0.09}Te_2$ at LT and RT. (c,d) Photoelectron spectra at $k_X=const.$ extracted at RT and LT for $E_B$ = 15 meV along the section "1" marked in (e,f). (i,j) Schematic representation of the $E$-$k$ dispersions of the valence and conduction bands of a type-II Weyl semimetal for RT and LT. The photon energy is 21.2 eV.

The schematic representation of the $E$-$k$ dispersion [see Figs. 4(i,j)] explains the appearance of the TSS in the inversion symmetry broken $T_d$ phase in contrast to the inversion symmetric 1$T'$ phase. The tilted type-II Weyl cones form at the intersection of conduction (CB) and valence bands (VB). The inversion symmetry breaking constrains the minimum number of Weyl nodes to four. The projected bulk electron (red) and hole (blue) pockets coexist and touch at the type-II Weyl nodes (shown as black dots). The TSS (green dotted lines), connect pairs of Weyl nodes, which are below and above $E_F$ [55].

Some of our photoemission results show that the band structure appears sharp at RT and blurred at temperatures below 170 K (see Supplementary Material Fig. S5). This is counterintuitive, because typically the valence bands appear sharper at low temperature due to a decreased electron-phonon scattering. The most likely reason of our observation is the occurrence of structural domains at low temperature as reported in Ref. [28], where the authors performed a spin-resolved laser ARPES study of 1$T'$-$MoTe_2$ and revealed two types of domains with different surface band dispersions. These domains have an opposite bulk polarity of the crystal structure. Ref. [28] also reports that Fermi arcs have non-equivalent shapes for both domains, connecting the same type of W1 nodes (domain B) or different pairs of W1 and W2 nodes (domain A), respectively. These experimental results were substantiated by slab calculations on the (00−1) and (001) surfaces, respectively. Therefore, the region of interest on the sample was adjusted on a single domain in order to avoid the overlay of several domains.

### c. Comparison of experiment and theory

From the XRD measurements, precise atom positions for the 1$T$` and the $T_d$ phase were derived. This experimentally determined structure was used for DFT calculations of the bulk band structure. Comparison of the calculated and experimental band structures shows good agreement (Fig. 5).

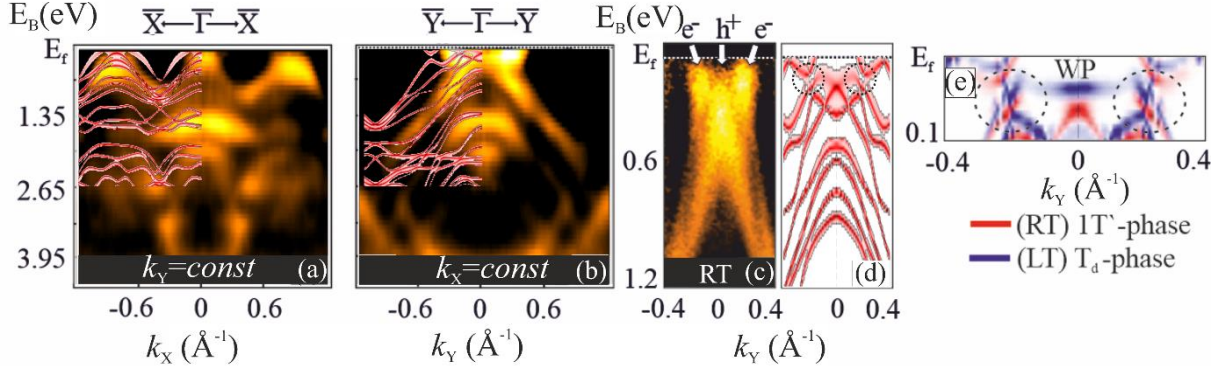

**Figure 5.** Comparison of calculated band structures for Mo$_{0.91}$W$_{0.09}$Te$_2$ based on experimental XRD data and experimental photoemission patterns for (a) $k_Y = 0$ and (b) $k_X = 0$. Magnified area from (c) experimental and (d) calculated data close to the Γ-point ($k_X = 0$). (e) Detail close to the Γ-point, where a Weyl point could appear at LT as marked with dashed circles. The red/blue color code indicates results for the 1T'- and $T_d$ -phase, respectively.

Since the XRD and ToF-MM measurements were performed on the same samples, the results of the DFT calculations are consistent and could explain the redistribution of intensity in the photoemission experiment at different temperatures (see Fig. 4). For example, Fig. 5(e) clearly shows how the valence bands shift to lower binding energies at LT (bands are marked in blue) and touch the conduction bands at points close to ±0.2 Å$^{-1}$. This result supports our explanation of a surface state appearing in the $T_d$ phase and agrees one to one with the intensity profiles [see Figs. 4(c,d)].

### IV. Raman Spectroscopy

To confirm that at low temperature the Mo$_{0.91}$W$_{0.09}$Te$_2$ sample measured with high-resolution photoemission spectroscopy is indeed in the $T_d$ phase, we performed temperature and polarization dependent Raman spectroscopy measurements. We used polarization dependent measurements to confirm the crystal structure and orientation of it. Thus, the *a*-axis of the crystal was parallel to the polarization of the incident light, *i.e.* "*xx*" polarization (see Fig. 6, where the *a*-axis corresponds to zero angle).

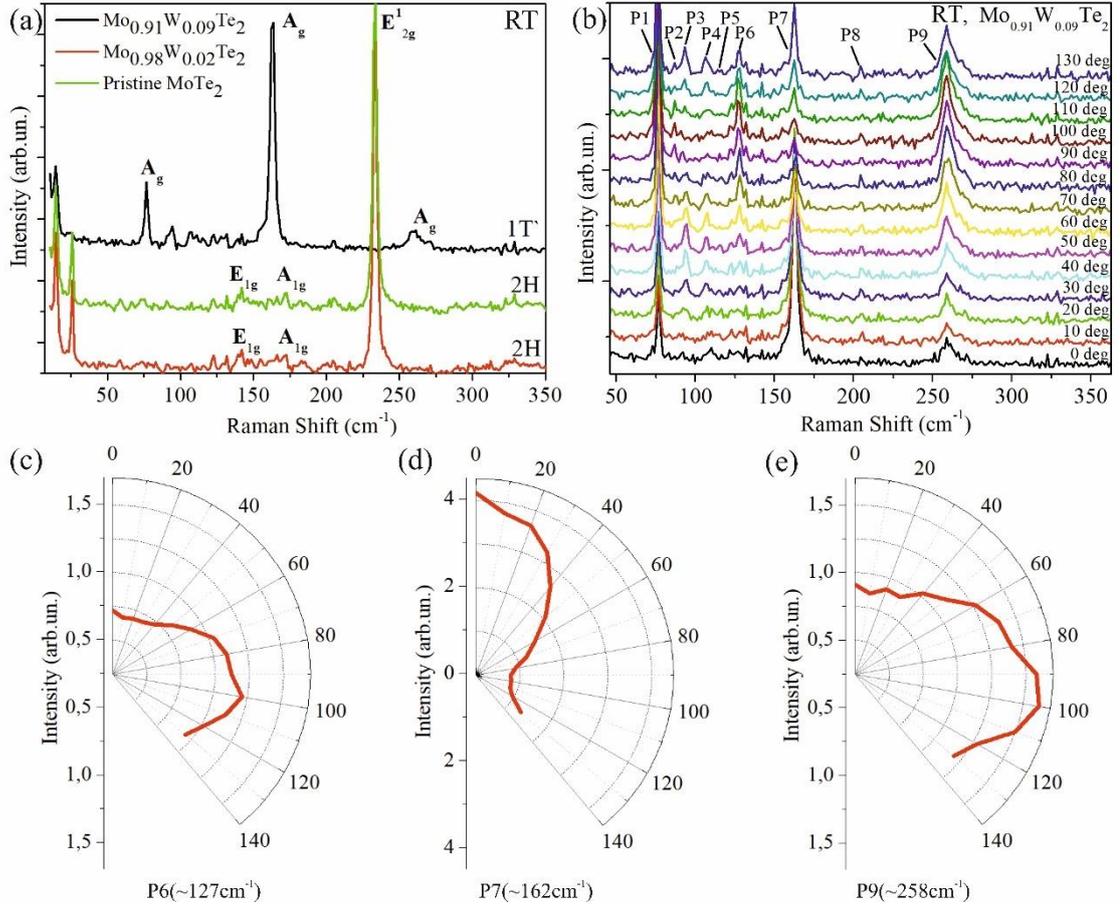

**Figure 6.** (a) Raman spectra for bulk $Mo_{0.91}W_{0.09}Te_2$ (1T'), $Mo_{0.98}W_{0.02}Te_2$ (2H), and $MoTe_2$ (2H) at room temperature (RT) in parallel polarization configuration. (b) Raman spectra for bulk $Mo_{0.91}W_{0.09}Te_2$ (1T') for different sample rotation angle. (c-e) Polarization dependence of representative Raman peaks measured in parallel polarization configuration.

The Raman spectra for bulk $Mo_{0.91}W_{0.09}Te_2$ comprise two characteristic peaks due to the $A_g$ and $B_g$ vibrational modes of the 1T'-$MoTe_2$ structure. In bulk crystals, the structural phase transition from the 1T' to the $T_d$ phase gives rise to remarkable Raman signatures. Fig. 6(a) compares the Raman spectra of $Mo_{1-x}W_xTe_2$ at RT in three different phases (x=0; 0.02; 0.09). The polarization dependent measurements allowed us to identify the symmetry of the detected modes and the crystalline orientation, through a direct comparison with data from the literature [56]. The measured modes show a periodic variation over 180° of azimuth angle with a two-lobed shape [Figs. 6(c-e)]. The modes (~127 cm$^{-1}$) and (~162 cm$^{-1}$) of $Mo_{0.91}W_{0.09}Te_2$ show an in-plane anisotropic Raman response similar to that of pristine 1T-$MoTe_2$ in Ref. [34,56].

The Raman selection rules restrict the observable modes: $A_g$ modes of the 1T phase (79, 109, 112, 129, 165, 250 and 260 cm$^{-1}$) are observed in parallel configurations along the crystal axes [(xx) or (yy)]; $B_g$ modes of 1T (93, 96 and 192 cm$^{-1}$) are observed in the cross-polarization [(xy) or (yx)]. This explains why P6 and P9 modes have their maximum intensity at the azimuthal angle 90° in the polar plots. In contrast, the P7 mode has its maximum at the azimuthal angle of 0°.

Raman spectroscopy measurements for the 2H and 1T phases agree with previously published results [56], [57]. The 1T and 2H phases have a common c-axis lattice constant, and, apparently, they have a similar interlayer distance. However, the 1T and $T_d$ phases have a different monolayer structure as compared to the 2$H$ phase, and thus the simple comparison of the overall interlayer distance hardly captures the difference in the interlayer coupling strength [57]. The interlayer atomic spacing between tellurium atoms (Te-Te distances) should be compared, as they play a major role in the total interlayer interaction. For example, some tellurium atoms in the 1$T'$ and $T_d$ phases have a much larger interlayer Te-Te distance compared to the 2$H$ phase (3.945 and 4.837 Å, respectively). This certainly leads to a weaker interlayer coupling compared to the 2$H$ phase.

The Raman spectra of the LT $T_d$ phase show distinct differences from 1$T'$: the modes at ~130 cm$^{-1}$ (P6) and ~257 cm$^{-1}$ (P9) are split into two peaks each. Since these two phases differ in the way the layers are stacked, the corresponding modes are likely interlayer modes. On the other hand, the difference in the crystal symmetry should lead to different selection rules for the two phases [57]. Both the 1$T'$ and the $T_d$ phase contain repeatable parts: 2 layers with 12 atoms in the unit cell, so each of the parts has a total of 36 vibrational modes. The 1$T'$ vibrational modes are decomposed into several modes $\Gamma_{bulk,1T'}$= 12$A_g$ + 6$B_g$ + 6$A_u$ + 12$B_u$ (where $A_g$, $B_g$ are Raman active modes, whereas $A_u$, $B_u$ are Raman-inactive). The $T_d$ vibrational modes are decomposed into $\Gamma_{bulk,Td}$= 12$A_1$ + 6$A_2$+ 6$B_1$ + 12$B_2$ (where all modes are active). Based on the Raman selection rules, the $A_g$ modes of the 1$T'$ phase and $A_1$ of the $T_d$ phase are observed in the parallel polarisation configuration along the crystalline axes, whereas the $B_g$ modes of the 1$T'$ phase and the $A_2$ modes of the $T_d$ phase are observed in cross polarization configuration. The comparison of the polarisation behaviour of the Raman peaks shows that P1, P4, P6, P8 are observed in the parallel polarisation as $A_g$ modes for the 1T' phase and P2, P3, P5 are observed in cross polarisation as $B_g$ modes for the 1T' phase. Thus, when the phase transition happens, the $A_g$ and $B_u$ modes (1T') transform into $A_1$ and $B_2$ ($T_d$) and for example the P$_{6a}$ peak at ~128 cm$^{-1}$ appears [Figs. 7(e,f)]. The phase where the peak splits into P$_{6a}$ and P$_{6b}$ is identified as the $T_d$ phase, and those with a single peak as the 1T' phase. Ref. [58] reports the evolution of the P$_6$ into the modes P$_{6A}$ and P$_{6B}$ in the heating and cooling cycles for Mo$_{0.8}$Re$_{0.2}$Te$_2$. Thus, the mode frequencies, line widths and integrated intensities of P$_{6A}$ and P$_{6B}$ exhibit a hysteresis at the phase transition (not only in resistivity). This is expected as the breaking of inversion symmetry leads to a first-order phase transition in 1T'-MoTe$_2$ that is accompanied by a complete renormalization of the phonons resulting in a hysteretic character of their parameters [34], [58], [59]. For example, in Ref. [58] the hysteresis in the intensities of modes P$_{6A}$ and P$_{6B}$ for (Mo,Re)Te$_2$ is compared with the hysteresis for pristine MoTe$_2$. It was shown that Re doping shifts the centre of the hysteresis loops by about 30 K (Re at the substitutional sites causes electron doping).

Not all active modes might be visible due to small scattering cross sections (depending on the exact atomic positions). Thus, peak intensities could be different for the two phases, even if the modes are allowed by the Raman selection rules. Some phonon modes of two-dimensional materials have a significant dependence on composition (peak position and width). For example, Ref. [59] reports that 2H-MoTe$_2$ has a shift of the $A_{1g}$ mode (~170cm$^{-1}$) of ~ 1 cm$^{-1}$ and an increase in the FWHM of ~ 1 cm$^{-1}$. Another possible origin of peak shifts is the variation of the local strain (relative sliding of adjacent layers for different phases). Ref. [60] reports Raman shifts of 1 cm$^{-1}$ and 4 cm$^{-1}$, respectively, for the $A_{1g}$ (~ 174 cm$^{-1}$) and $E^1_{2g}$ (~ 236 cm$^{-1}$) modes in the range from 80 to 300 K. Raman shifts for both modes decrease linearly with increasing temperature. The authors attribute this frequency shift to lattice thermal expansion and decay into lower energy phonons.

For monoclinic or orthorhombic TMDCs, one can observe phonons that can be divided into two groups: 1/3 of the modes are *z* modes vibrating along the direction of the zigzag transition-metal atomic chain and 2/3 are *m* modes that vibrate in a mirror plane (perpendicular to the zigzag chain) [25]. When the incident and scattered light are polarised perpendicular to each other, all *z* and *m* modes exhibit 4-fold symmetry. In the case of parallel polarisation, the angular dependence for the *m* modes is very sensitive to specific lattice vibrations [25].

It is intriguing that the P$_{6B}$ and P$_{9B}$ modes (but not the P$_{6A}$ and P$_{9A}$ modes) vanish at high temperatures [see Fig. 7(c-f)]. The reason for this effect is still unclear. The integration of W into substitutional sites is a complicated process with charge redistribution and local distortions in the crystal that can lead to a modification of the phonon spectra.

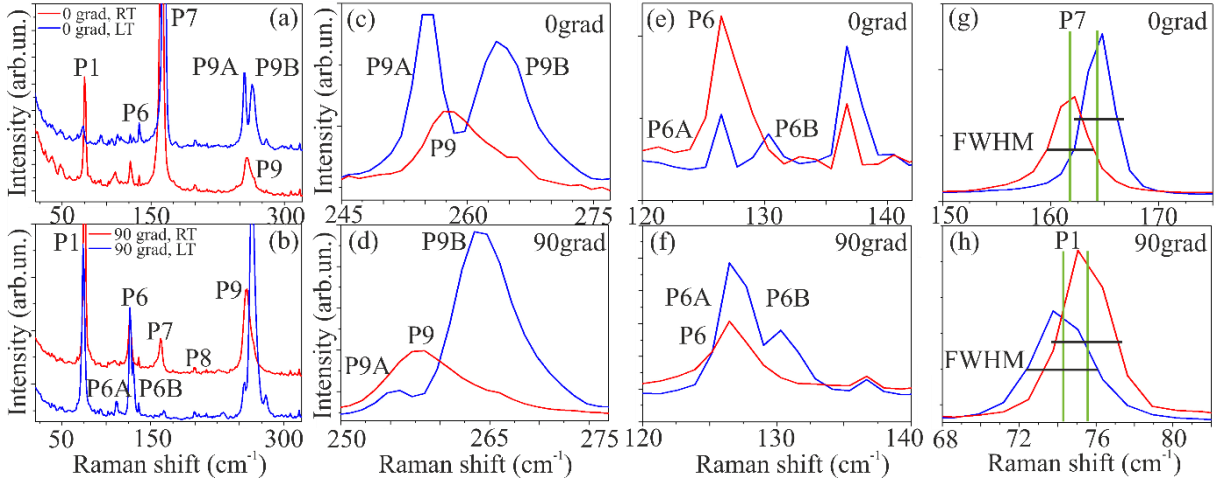

**Figure 7.** Raman spectra measured at RT and LT. (a,b) Single symmetric modes P6 and P9 in 1T` phase. The T$_d$ phase shows splitting into two modes P6A, P6B and P9A, P9B because of lack of the inversion symmetry at 0° (c,e) and 90° (d,f) azimuthal angle. (g,h) Shift and FWHM of P7 and P1 modes, which change at RT and LT. Green lines are a guide-to-the-eye, quantifying the shift of the P7 and P1 modes.

It is also interesting that the P7 mode shows a redshift (to lower values) and an increase in linewidth with increasing temperature. This behaviour can be explained by the anharmonic approximation for phonons (including thermal expansion contributions). The frequencies and linewidths are described by cubic anharmonic equations [58], [61].

As an exception, the P1 mode at 75 cm$^{-1}$ increases in frequency with increasing in contrast to the usual frequency decrease expected for anharmonic effects and observed for the other phonon modes. This might be a hint that the P1 mode is affected by electron-phonon coupling. On the other hand, the P1 mode does not appear asymmetric in shape, which would be a direct fingerprint of electron-phonon coupling.

The calculated energy difference between the 1*T* phase and the *T*$_d$ phase is only a few meV per unit cell [62]. As the temperature decreases, the total energy of the T$_d$ phase decreases faster than that of the 1*T* phase, leading to the 1*T* to *T*$_d$ phase transition [57]. This phase transition involves a slide along the "*y*" direction. The interlayer coupling constant along the *y* direction is very small, thus even a slight perturbation by strain can trigger the phase transition [57]. As the sample becomes thinner, the unbound surface layers (top and bottom) become significant, and the difference in the total energy becomes even smaller, leading to the coexistence of both phases. Compared to the 2*H* phase, the 1*T* and *T*$_d$ phases have much weaker interlayer coupling along in-plane and out-of-plane directions.

The authors of Ref. [62] theoretically predicted the stabilisation of the *T*$_d$ or 1*T* phases by electron or hole doping. Furthermore, hole doping can stabilise the 1*T* phase even in a wide temperature range (80-400 K). In general, there is a competition between the following sources of electron or hole doping: Electron doping is caused by Te deficiencies and structural defects due to exfoliation, and by partially replacing Mo atoms by W atoms. Hole doping is due to the exposure to ambient pressure and the chemical reaction with oxygen. In Ref. [63] it has been shown that the topological phase transition in bulk 1$T$'-MoTe$_2$ is related to electron-phonon coupling. Indeed, the latter is a key mechanism for changing the main topological features in the electronic structure of MoTe$_2$. The $^2A_1$ mode (~76 cm$^{-1}$) exhibits an unusual softening of the frequency broadening in FWHM with temperature decrease, in contrast to other modes that show normal hardening and narrowing due to anharmonic phonon coupling. In general, phonon softening can occur for three reasons: structural phase transition, lattice expansion, and phonon coupling with other quasiparticles (see Supplementary Material of [63]).

## V. XRD and XPD measurements

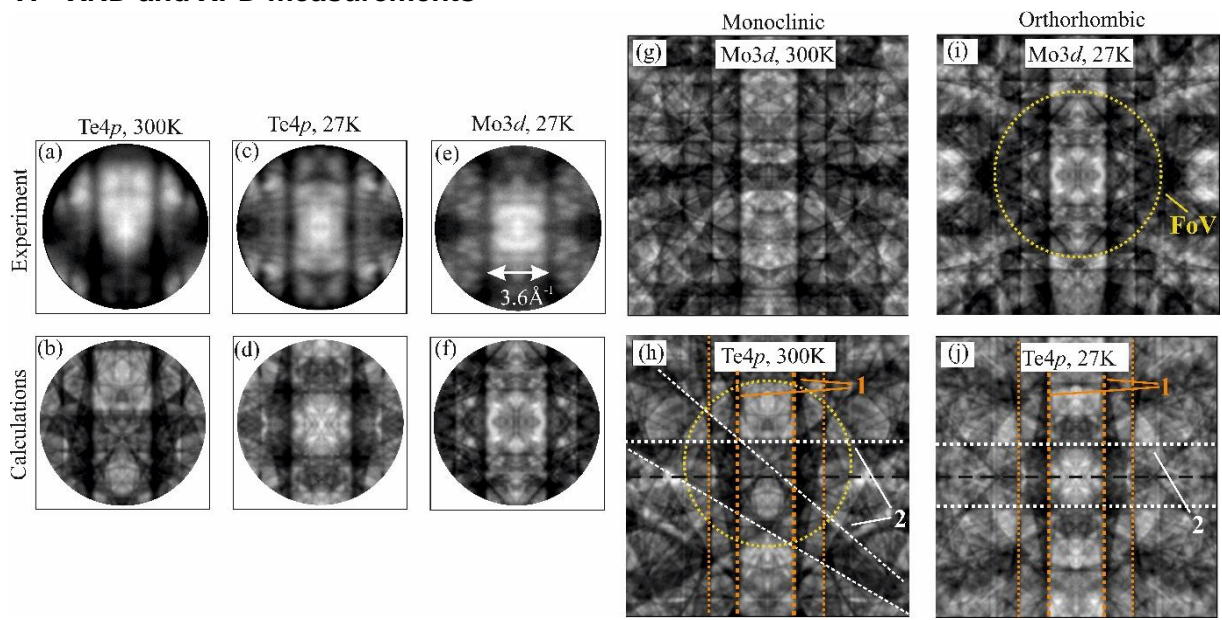

**Figure 8.** Measured and calculated XPD patterns. (a,c,e) Measured XPD patterns for Mo$_{0.91}$W$_{0.09}$Te$_2$ at hv = 5120 eV for the Te 4$p$ core level at (a) RT and (c) LT; and for the Mo 3$d$ core level at LT (e). (b,d,f) Corresponding calculated XPD patterns for Mo$_{0.91}$W$_{0.09}$Te$_2$ using the Bloch-wave approach. (g,i) Calculated XPD patterns for the Mo 3$d$ core level for the 1$T$'- and $T_d$ phase, respectively. (h,j) Similar calculated data for the Te 4$p$ core level for 1$T$' and $T_d$ –Mo$_{0.91}$W$_{0.09}$Te$_2$. The orthorhombic phase shows up/down (2) and left/right (1) symmetry in XPD patterns and the monoclinic phase only left/right (see orange and white dashed lines). The yellow circle represents the field of view which was used in the experiment.

XRD measurements indicate the monoclinic 1$T$'-MoTe$_2$ structure in the $P2_1/m$ space group with lattice parameters of $a$ = 6.329 Å, $b$ = 3.479 Å, $c$ = 13.933 Å and $β$ = 93.795° (unit-cell volume of 306.126 Å$^3$) at 300K and transition to the orthorhombic $T_d$ structure in the Pmn2$_1$ space group with parameters of $a$ = 6.311 Å, $b$ = 3.469 Å, $c$ = 13.894 Å and $β$ = 90.0° (unit-cell volume of 304.181 Å$^3$) upon cooling. XRD measurements yielded the precise positions of atoms in the cell (a complete structural model) for the 1$T$' and $T_d$ phases (see Supplementary material, T1), that was used for theoretical calculations of XPD for Mo and Te core levels. A comparison of the experimental and calculated XPD patterns is presented in Fig. 8. The XPD calculation was performed using the Bloch-wave approach, which showed a one-to-one agreement with the experiment [64], [65]. In the case of Mo$_{0.91}$W$_{0.09}$Te$_2$, the calculated XPD patterns show significant differences for RT and LT. This is further evidence that Kikuchi diffraction is very sensitive to structural changes [66]. Thus, at the phase transition from 1$T$' to $T_d$ the angle β changes only by 3.8°, the lattice parameters *a* and *b* change by less than 0.02 Å, and *c* by less than 0.04 Å.

In comparison to the 2*H* phase, the 1*T'* and $T_d$ phases are structurally comparable and belong to a similarly distorted 1T phase. The structural distinction between 1*T'* and $T_d$ can only be observed for few-layer crystals [1]. Few successful phase identifications have been reported, *e.g.,* by scanning transmission electron microscopy (STEM) with atomic resolution [67]. However, we indeed observe these small differences with XPD as shown in Fig. 8.

Calculated Kikuchi patterns, including projections of lattice planes (hkl) and Kikuchi bands (extended from planes by ± Bragg angles), on the base of experimental XRD data are presented in the Supplementary Material (Fig. S8). Some of those Kikuchi bands are shown as dashed lines in Figs. 8(g-j) as a guide-to-the-eye to structural differences between 1*T'* and $T_d$ phases.

Experimental XPD patterns are in a good agreement with calculated ones and show 2-fold symmetry for both phases 1*T'* and $T_d$. It is inspiring to note, however, that the XPD patterns show left/right and up/down symmetry for measurements at 20 K, and only left/right (up/down symmetry is broken) for measurements at 300 K (see wite and yellow lines in Fig. 8,h-j). This behaviour is counterintuitive, because one would expect the same effect to appear in the XPD patterns for the LT phase with inversion symmetry breaking. However, the RT phase is a monoclinic phase with layers shifted with respect to each other (β=93.8°). This shift breaks the up/down symmetry, and this is not the case for the orthorhombic phase at 20 K. Thus, XPD patterns are very sensitive to small changes in crystal structure that happen due to the phase transition.

## VI. Electronic Transport

Transport measurements of the 2*H*, 1*T'* and $T_d$ phases of $Mo_{1-x}W_xTe_2$ were performed under varying magnetic field and strain. Experimental details are given in the Supplementary Material (Figs. S7-S9).

The temperature dependence of the electrical resistivity (ρ) of $Mo_{1-x}W_xTe_2$ (*x*=0; 0.02; 0.09) was measured down to a minimum temperature of 1.4 K. Doped samples (with 2 and 9% of W) show a hysteresis of the temperature dependent resistivity [see Fig. 9(a)]. It is intriguing that $Mo_{0.98}W_{0.02}Te_2$ exhibits a broad thermal hysteresis of ρ at $T_S$=240 K (where $T_S$ is a phase transition temperature), but Raman spectroscopy measurements show that the 2*H* phase persists at this temperature [see Fig. 6(a)]. Therefore, the resistivity curves during heating and cooling should be similar to those of the pristine sample. The hysteresis could be an indication that at 2% W doping a coexistence of the two phases 2*H* and 1*T'* takes place. This is consistent with other reports [35], [34]. At larger doping concentration (9% of W) the transition temperature is shifted to higher values ($T_S$= 265 K) and makes the hysteresis loop smaller because the 1*T'*-phase dominates in this case. This also agrees with theoretical predictions, where a high doping concentration leads to a stabilization of the orthorhombic $T_d$ phase at RT [9]. The first-order structural phase transition from the 1T` to the $T_d$ phase is associated with a thermal hysteresis.

The residual resistance ratio (RRR), ρ(*T*=300K)/ρ(*T*=0 K), gives an estimate of the sample quality as the *T*=0 resistance occurs through elastic scattering on defects or impurities. Our measurements showed that while pristine 2H-$MoTe_2$ has a giant RRR (of 1200), $Mo_{1-x}W_xTe_2$ (0.02; 0.09) values are of only 4 and 8, respectively. This difference in RRR could be related to the different methods of samples growing. For example, the doped samples were grown using a fast quenching procedure. In Ref. [63] the authors summarize temperature dependent RRR phase diagrams for $MoTe_2$ samples obtained by different synthesis methods and propose a temperature – RRR phase diagram: the phase at higher temperatures is in one topological state (TP I with 8 Weyl-points), while the phase in the high RRR (>100) and low-temperature (<75 K) region is in another topological state (TP II with 4 Weyl-points). Upon cooling MoTe$_2$ with high-RRR a transition from the TP I to TP II Weyl semimetal state occurs. In contrast, the MoTe$_2$ with low-RRR remains in the TP I state throughout the entire temperature range. In our case, Mo$_{0.91}$W$_{0.09}$Te$_2$, that is in the topological T$_d$ phase at LT, has a residual resistance of RRR = 8. This means that the topological state should be stable over a wide temperature range. Yet, according to the DFT calculations we expect the occurrence of 4 Weyl points instead of 8.

Our temperature-dependent ρ(T) measurements show that as the doping concentration increases from 2% to 9%, the resistivity decreases significantly at low temperatures [see Supplementary Material Fig. S6(a)]. It can be assumed that with even higher doping concentration, the superconductor transition temperature $T_c$ could be increased, and the phase transition would occur faster and even at room temperature ($T_S$ shifted to higher values). The increase of $T_C$ for Mo$_{0.91}$W$_{0.09}$Te$_2$ in comparison to Mo$_{0.98}$W$_{0.02}$Te$_2$ is explained by the increase of the electronic density of states at the Fermi level [see Fig. 3(c)]. In contrast to the doped samples, pristine 2H-MoTe$_2$ shows a very high and rapidly growing ρ at low temperatures [see Supplementary Material, Fig. S6(b)].

The $T_d$ phase exhibits a large magnetoresistance effect in the Weyl semimetal state, a state of matter in which collective excitations, known as Weyl fermions, can exist. The Fermi surface is highly sensitive to even small changes in the lattice constants, which can be induced by strain or as a function of temperature [28].

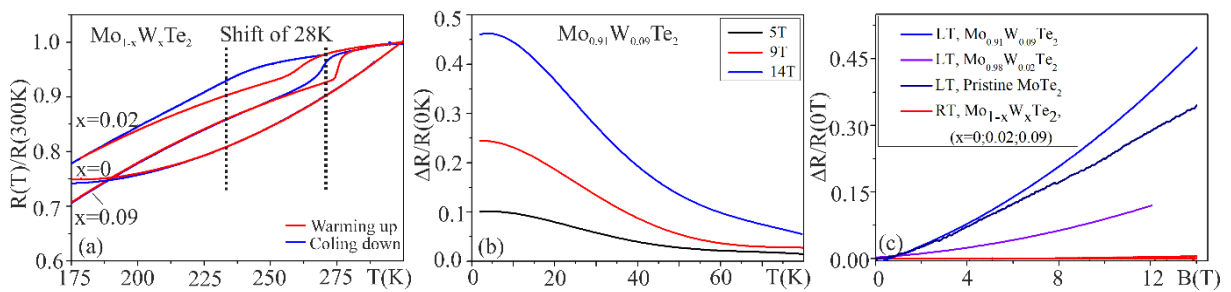

**Figure 9.** (a) Resistivity of Mo$_{1-x}$W$_x$Te$_2$ (x=0; 0.02; 0.09). (b) Magnetoresistance of Mo$_{0.91}$W$_{0.09}$Te$_2$ at different magnetic fields. (c) Magnetoresistance of Mo$_{1-x}$W$_x$Te$_2$ (x=0; 0.02; 0.09) at different temperatures. The magnetic field was applied along the crystallographic c-axis.

The enormous magnetoresistance [see Fig. 9(c)] is explained by the perfect balance between the electron and hole populations and the field-induced enhancement of electron backscattering, as explained for Dirac semimetals [68]. According to Ref. [68], such backscattering is forbidden by the symmetry at zero field and lifted by the magnetic field. The theoretical band structure of $T_d$-MoTe$_2$ reveals a non-trivial orbital and spin texture of the pockets. The spin texture was confirmed experimentally in the related compound WTe$_2$ [69], where the circular dichroism in photoemission suggests spin polarised Fermi surfaces.

In semimetals such as 1T'-MoTe$_2$, the temperature-dependent resistivity can be expressed as $ρ = a+bT+cT^2$, where the first term originates from electron-defect scattering, the second term from electron-phonon scattering, and the third term from electron-electron scattering. At low temperatures, where phonon scattering is negligible, the ρ exhibits Fermi liquid behaviour, while at higher temperatures, the resistivity increases almost linearly with temperature. The carrier mobility of the both 2H and 1T' phases decreases with temperature.

Qi *et al.* [7] showed superconductivity at ambient pressure at $T_c = 0.10$ K in bulk $T_d$-MoTe$_2$. $T_c$ could be strongly enhanced by applying external pressure. Bulk $T_d$-MoTe$_2$ exhibits topological s+- superconductivity and type-II WS, while monolayer $T_d$-MoTe$_2$ is predicted to be a topological insulator [1].

The DFT calculations in Ref. [70] show significant *d-p* orbital mixing at the Fermi level, indicating non-trivial orbital pseudospin texture. The electron and hole pockets exhibit the chiral nature as Weyl points. Carrier scattering between pockets of opposite chirality (on opposite sides of the $\Gamma$-point) is suppressed due to the opposite sign of the Berry phases of the time reversal-related scattering paths. In the presence of magnetic field, the broken time-reversal symmetry leads to cancellation of the Berry phases and an enhanced inter-pocket backscattering [70]. Since the orbitals are directly coupled to the lattice, the application of strain to a Weyl semimetal is expected to alter the orbital texture of the pockets and hence, the inter-pocket electron scattering. Since Dirac cones are tilted along the "*a*" direction, different effects of tensile strain applied along "*a*" and "*b*" might be expected. The authors of Ref. [70] propose that the strain-induced fields, unlike the real magnetic field, preserve time-reversal symmetry. This explains the small changes in strain-dependent resistivity at zero magnetic field. Thus, large strain-induced magnetoresistance (SMR) originates from a non-trivial interplay of magnetic- and strain- induced fields.

Spintronic devices exhibit a variety of strain-induced effects, including modulation of the band gap, carrier mobility, and light absorption. Manipulation of the band gap by applying the strain can lead to significant changes in physical properties. According to Ref. [70], the magnetoresistance in MoTe$_2$ is enhanced by 30 % at low temperatures and high magnetic fields (9 T) under uniaxial strain ($\varepsilon=\Delta L/L=4.8\cdot10^{-4}$) along the "*a*" crystallographic direction (due to in-plane anisotropy), and reduced by the same amount when $\varepsilon \parallel b$. This is corroborated by calculations that show a significant change in the Fermi surface (non-trivial spin-texture at the Weyl point) under tensile strain [70].

Here we present measurements that were made with both electric current and tensile strain applied along the *a* (long one) crystallographic axis. For experimental details see Supplementary Material (Fig. S7).

The temperature dependence of the relative change of $\rho$ as a function of applied strain (strain induced resistance) is presented in Fig. 10. It is clearly seen that at temperatures below the phase transition from $1T$ to $T_d$ the strain induced resistance increases significantly, as the magnetoresistance does. Thus, measurements at 300 K and 200 K showed no change in resistance with applied strain. Our measurements are in agreement with transport measurements for pristine $1T$-MoTe$_2$ [70], where it was shown that when $\varepsilon \parallel b$, a clear decrease in $\rho$ with increasing strain is observed. The lower the temperature, the larger is this decrease. When $\varepsilon \parallel a$, the effect is reversed: the resistivity is enhanced, and the longitudinal strain resistance is larger at lower temperatures. In Ref. [70] it was shown that the structural transition temperature is shifted when applying strain. Thus, during cooling, the temperature interval of the transition region between the two phases $1T$ and $T_d$ is stretched when $\varepsilon \parallel b$ and compressed when $\varepsilon \parallel a$.

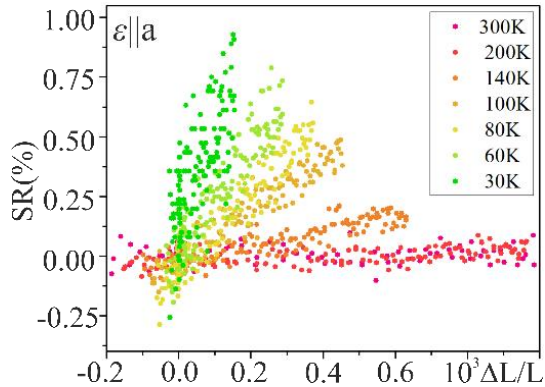

**Figure 10.** The temperature dependence of the resistivity as a function of applied strain (SR). Measurements were done during cooling down to 5 K at zero magnetic field.

Reversible switching between the metallic 1*T* phase and the semiconducting 2*H* phase by strain has been shown for MoTe$_2$ films in a field-effect transistor [71]. The calculation of the band structure under strain predicts modifications in the band dispersion near the Γ-point and along the Y → Γ direction, allowing a unique type of band manipulation via strain application [70]. When the strain is applied along the *a*-axis, the DOS at the Fermi level increases due to the shift of bands, and vice versa when strain is applied along the *b*-axis. The Fermi level has a more pronounced response to strain along the *b*-direction. The opposite trends of the DOS variation at different strain values are consistent with the observed strain-dependent resistivity effects.

## VII. SUMMARY

We have carried out a comprehensive experimental study of the electronic and structural phase transition in the bulk type-II Weyl semimetal Mo$_{1-x}$W$_x$Te$_2$. We have found that at a doping concentration of about 9% of W (*x*=0.09), the system exhibits a metastable monoclinic 1*T* phase at room temperature and an orthorhombic $T_d$ phase with broken inversion symmetry below 150 K. Changes in the electronic structure due to the phase transition at different doping concentrations were directly detected by angle-resolved photoemission measurements (including XPS and XPD), and changes in the crystal structure were detected by angle-resolved Raman spectroscopy. We have demonstrated the existence of surface states in the $T_d$ phase and associated them with the topological Fermi arcs. The experimental results show good agreement with DFT and Bloch wave calculations of the XPD patterns for both 1*T* and $T_d$ phases, giving indirect evidence that the observed surface states are topological Fermi arcs.

Transport measurements indicate a shift of the phase transition temperature ($T_s$) to higher values with increasing doping concentration. The low temperature orthorhombic $T_d$ phase exhibits a high magnetoresistance (MR) of up to 50% and strain induced resistivity (SR) of up to 90% when uniaxially strained along the *a*-axis.

Our results demonstrate that the type-II Weyl semimetal Mo$_{1-x}$W$_x$Te$_2$ is a promising material for investigating the interplay between elastic deformations and electronic properties for TMDCs. It also offers the interesting possibility of combining the ability to stimulate reversible phase changes between topological and trivial phases with a large energy splitting of WPs.

**Author Contributions**
The photoemission experiments in the laboratory and at the synchrotron facility were conducted by OF, MK, LO, DV, DK, SC, OT, YL. FKD and KR provided the CVT-grown single crystals of Mo$_{1-x}$TW$_x$Te$_2$. SMS, OF performed Raman scattering measurements. MF, OF did an experiment to investigate transport properties. MM carried out X-ray diffraction measurements. PR performed DFT calculations. XPD calculations based on the Bloch wave approach was implemented by AW and OF. The study was conceived and designed by OF, HJE, GS. The collaboration among the authors was organised and managed by OF. All authors discussed the results, edited, commented, and agreed on the manuscript.

**Competing interests**
The authors declare no competing interests.

**Acknowledgements**
The authors are grateful to Yong Sheng Zhao (DESY, AP-STAR Shanghai) for assistance in preparing samples for transport measurements.
This work was funded by the Deutsche Forschungsgemeinschaft (DFG, German Research Foundation) in the frame Individual Grants of FE2281/1-1 – 516525057 and SFB TRR228, projects B04 and B03 as well as Bundesministerium für Bildung und Forschung BMBF (grants 05K22UM1 and 05K22UM4).
PR and YuM acknowledge the support of the Joint Lab Virtual Materials Design (JL-VMD) and are grateful for computing time granted by the JARA Vergabegremium and provided on the JARA Partition part of the supercomputer CLAIX at RWTH Aachen University (project number jara0191). S.M.S. acknowledges funding by the DFG-Projektnummer 441231589. This work was furthermore funded by the DFG under Germany's Excellence Strategy – Cluster of Excellence Matter and Light for Quantum Computing (ML4Q) EXC 2004/1 – 390534769, and PR thanks the Bavarian Ministry of Economic Affairs, Regional Development and Energy for financial support within the High-Tech Agenda Project "Bausteine für das Quantencomputing auf Basis topologischer Materialien mit experimentellen und theoretischen Ansätzen".
AW has been supported by the Polish National Science Centre (NCN), grant number 2020/37/B/ST5/03669.
MF acknowledges funding from the Alexander von Humboldt foundation and the YIG preparation program of the Karlsruhe Institute of Technology.

*Data Availability*
The data of this study are available from the corresponding author upon request.
The calculated data was generated using the JuKKR code [42] via the AiiDA-KKR plugin [71], [73] to the AiiDA infrastructure and are publicly available on the materials cloud archive [74,75].

*Code Availability*
The source codes of the AiiDA-KKR plugin [73], [72] and the JuKKR code [42] are published as open-source software under the MIT licence at [73] and [42], respectively.

## Supplementary Material

*Sample characterisation*

XRD experimental data:

| Phase | Unit cell volume (Å³) | Lattice parameters (Å) | | | |
|---|---|---|---|---|---|
| | | a | b | c | β |
| Td | 304.181 | 6.311 | 3.469 | 13.894 | 90.000 |
| 1T' | 306.126 | 6.329 | 3.479 | 13.933 | 93.795 |
| Atoms position | | | | | |
| Phase | Element | x | y | z | Occ. |
| Td | Mo1 | 0.895 | 0.500 | 0.010 | 0.943 |
| | Mo2 | 0.530 | 0.000 | 0.996 | 0.870 |
| | Te1 | 0.285 | 0.500 | 0.108 | 1.000 |
| | Te2 | 0.790 | 0.000 | 0.151 | 1.000 |
| | Te3 | 0.362 | 0.000 | 0.355 | 1.000 |
| | Te4 | 0.857 | 0.500 | 0.398 | 1.000 |
| | W1 | 0.895 | 0.500 | 0.010 | 0.058 |
| | W2 | 0.530 | 0.000 | 0.996 | 0.130 |
| Phase | Element | x | y | z | Occ. |
| 1T' | Mo1 | 0.184 | 0.250 | 0.008 | 0.943 |
| | Mo2 | 0.681 | 0.250 | 0.494 | 0.870 |
| | Te1 | 0.588 | 0.250 | 0.103 | 1.000 |
| | Te2 | 0.903 | 0.250 | 0.852 | 1.000 |
| | Te3 | 0.444 | 0.250 | 0.648 | 1.000 |
| | Te4 | 0.057 | 0.250 | 0.396 | 1.000 |
| | W1 | 0.184 | 0.250 | 0.008 | 0.058 |
| | W2 | 0.680 | 0.250 | 0.494 | 0.130 |

Table **T1**. Structural parameters for 1T'- and $T_d$-$MoTe_2$ from experimental XRD measurements

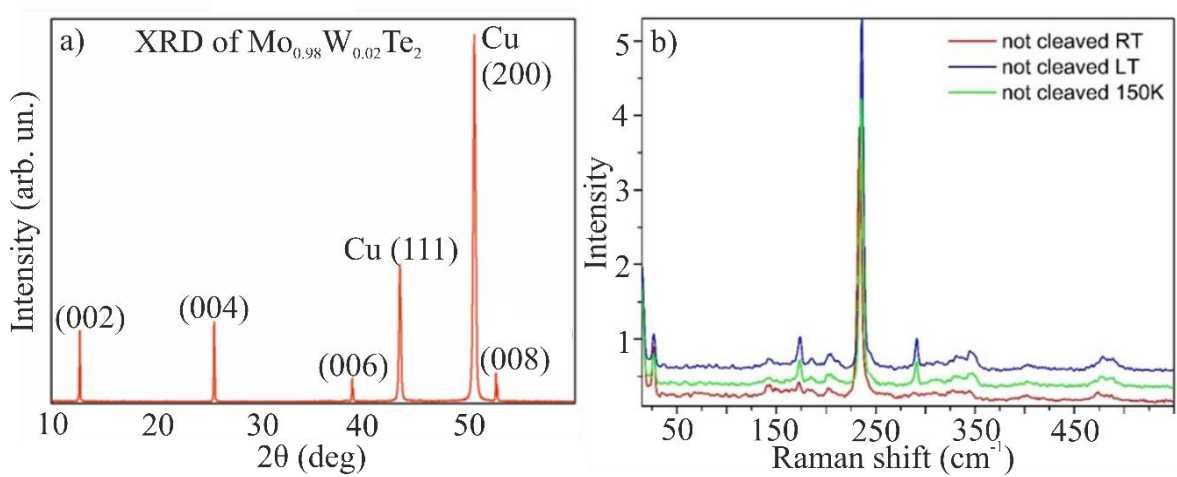

**Figure S1.** (a) XRD measurements of 2H-$Mo_{0.98}W_{0.02}Te2$ (a); Raman measurements for pristine $MoTe_2$ (2H-phase) in case of not cleaved sample at different temperatures: RT, 150 K and LT (25K)

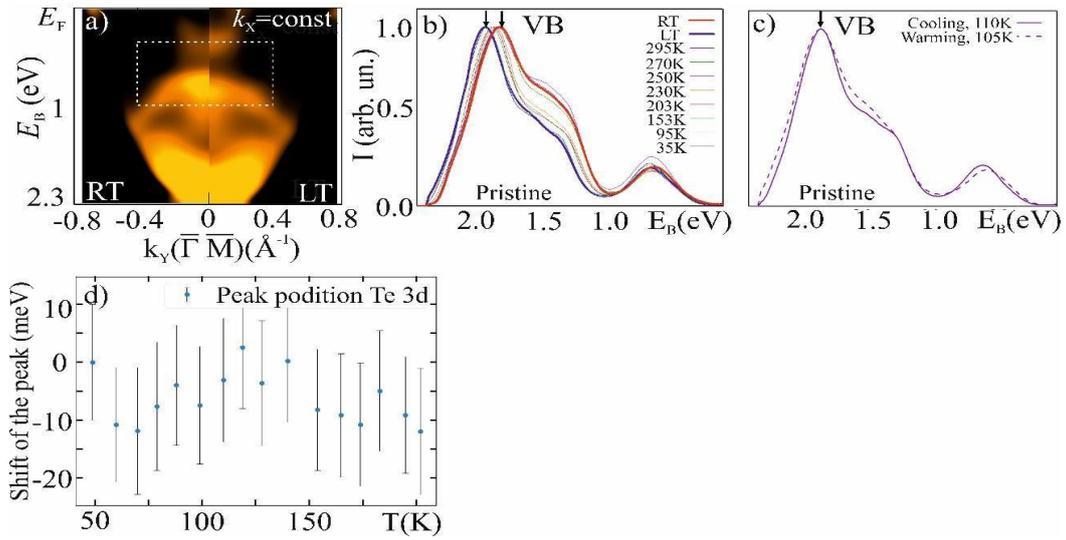

**Figure S2.** $E_B$-versus-$k_{||}$ section through 3D data array for pristine 2H-MoTe$_2$ at RT and LT (a); ToF spectra for VB measured during cooling (reversible during warming up) for pristine sample (b,c); ToF spectrum for core-level Te-3d$_{5/2}$ measured at different temperatures for Mo$_{0.91}$W$_{0.09}$Te$_2$ (d).

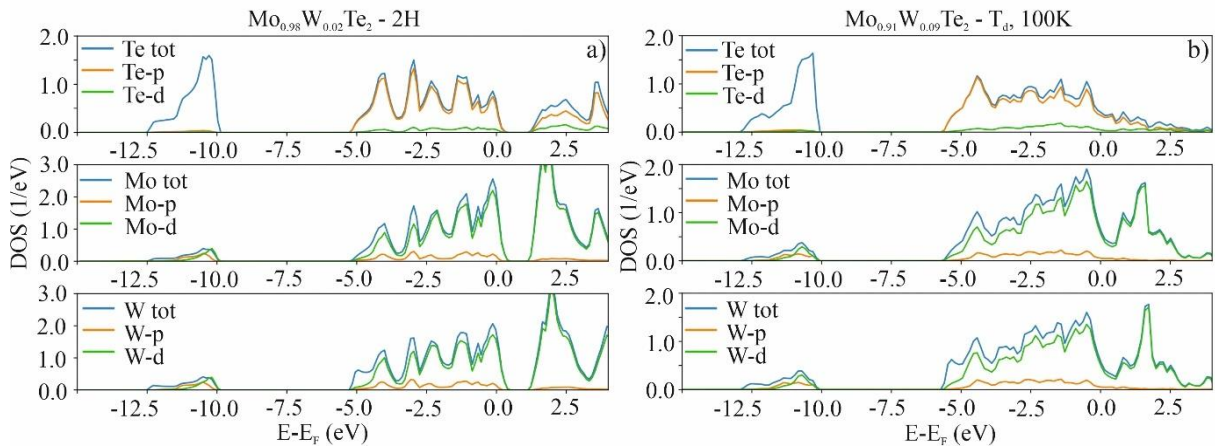

**Figure S3.** Calculated orbital projected DOS for Mo$_{0.98}$W$_{0.02}$Te$_2$ (a) and Mo$_{0.87}$W$_{0.12}$Te$_2$ (b) for Te4p, Mo5d and W5d.

Atoms position from XRD measurements were used also for DFT calculations of the bulk band structure (Fig.S4, a-b). Calculations were performed for RT (1T`-phase) and LT (T$_d$ -phase). Comparison of those two phases is presented in Fig.S4,c. Detailed sections for $k_Y$ = *const* and $k_X$ = *const* sections are presented in Fig.S4,d-e. In Fig.S9,e area, where CB contacts with VB is visible around 0.2 Å$^{-1}$ (see dashed line). In this area Weyl points appear and Fermi arcs are possible.

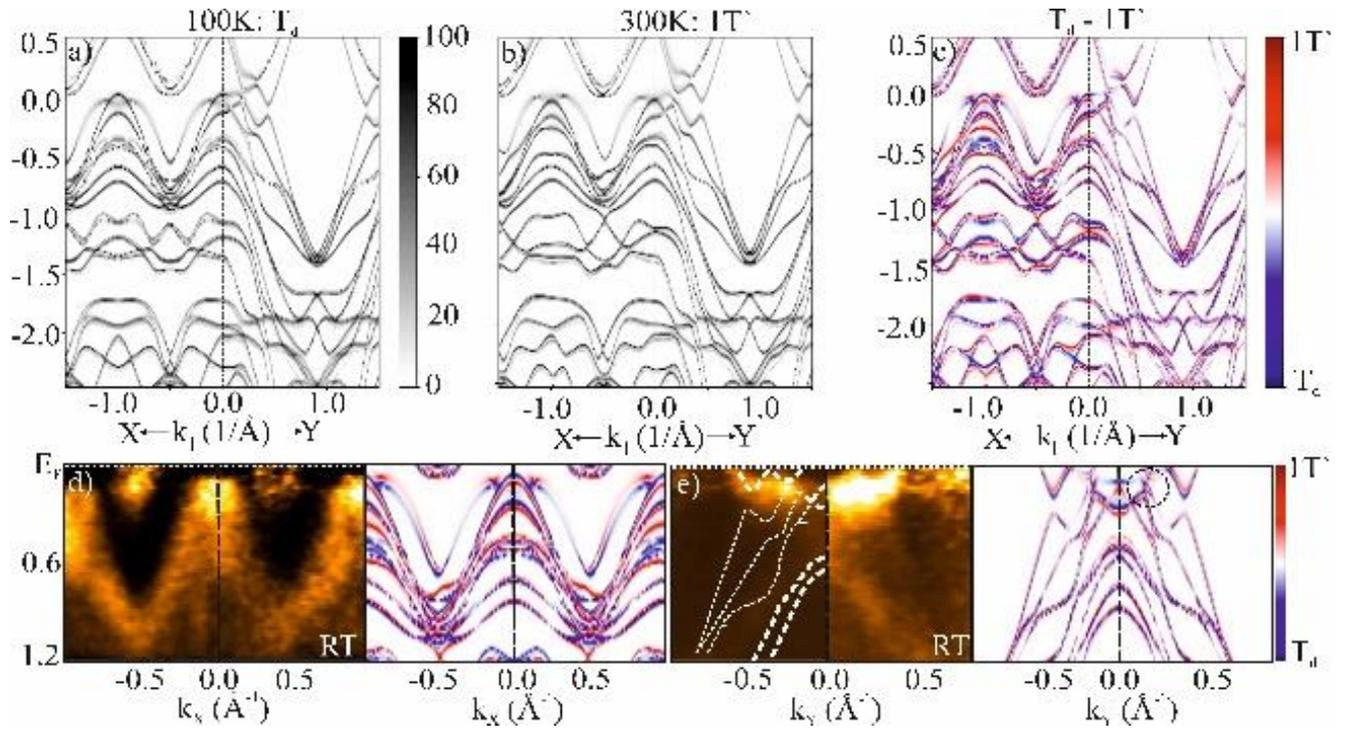

**Figure S4.** DFT calculations of bulk band structure for 1T`- and $T_d$ phases (a-c). Blue-red colour code corresponds to different temperatures: red – RT (1T`-phase), blue – LT ($T_d$ - phase); comparison of the calculated band structure and experimental photoemission images for $k_Y = 0$ (d) and $k_x = 0$ (e) sections.

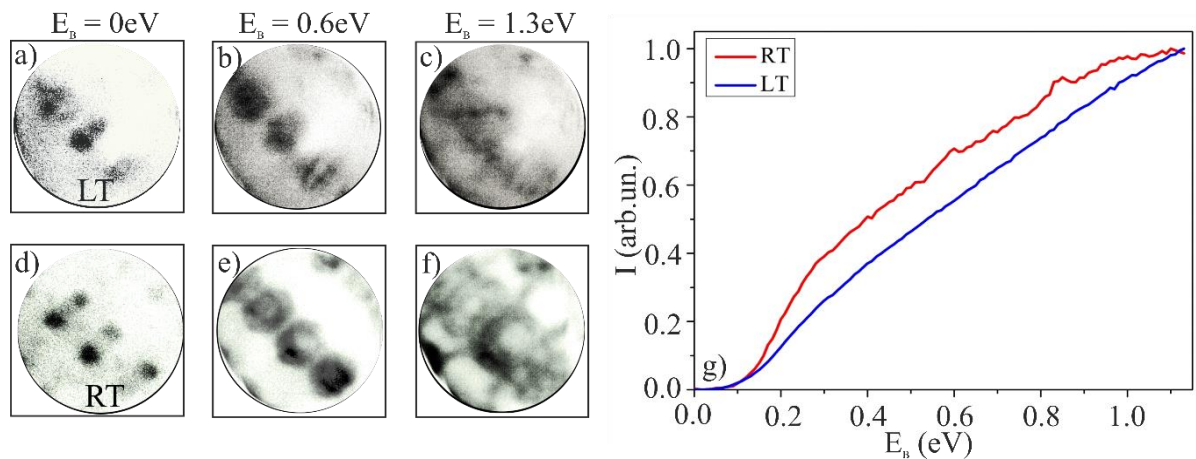

**Figure S5.** Constant energy sections through 3D data array as measured (without treatment) for LT (a-c) and RT (d-f); Comparison of ToF spectra measured at LT and RT (g).

Measurements were performed at different temperatures during cooling down and warming up. Photoemission patterns became diffuse during cooling down at 250 K and became again sharp during warming up. This effect is counterintuitive and could be explained by measurements at two domains with opposite bulk polarization at LT ($T_d$-phase with inversion symmetry breaking).

*Theoretical calculations*

XRD measurements gave an information about atoms position. Those data were used for calculations based on the Bloch-wave approach by A. Winkelmann. Theoretical Kikuchi pattern were obtained for 1T`- and $T_d$ – phases. Projections of lattice planes and extended (from planes by ± Bragg angles) Kikuchi bands reveal significant changes for different temperatures, namely absence of pattern symmetry. Thus, central Kikuchi bands are not symmetric up/down due to the cell distortion of monoclinic phase.

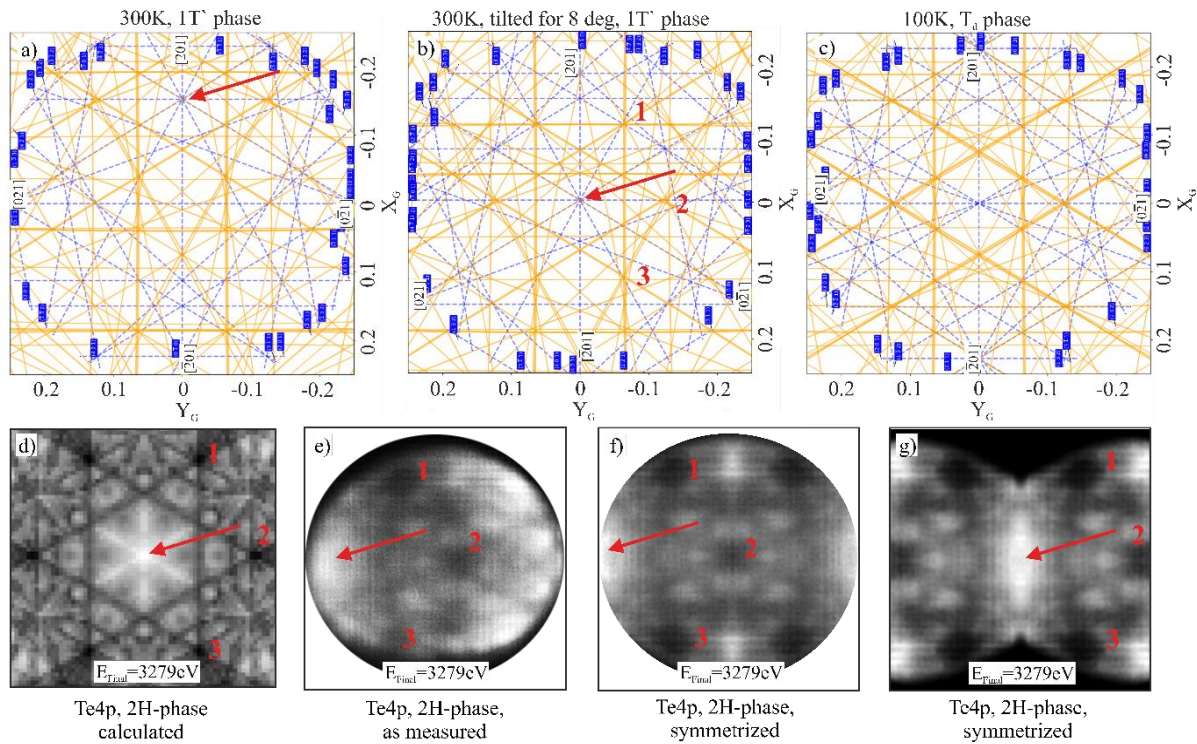

**Figure S6.** Theoretical Kikuchi pattern calculated on the base of Bloch-wave approach and experimental XRD data for $Mo_{0.91}W_{0.09}Te_2$ at 300 K (1T`-phase) (a); same but tilted for 8.5 deg (b); at 100 K ($T_d$-phase); calculated (d) and measured (e) XPD patterns for Te4p, 2H-$MoTe_2$ at hv=3340eV; measured XPD pattern symmetrized relatively to the center of the image (f) and relatively to the center of main Kikuchi bands marked with red arrow (g).

Theoretical Kikuchi pattern of $Mo_{0.91}W_{0.09}Te_2$ were calculated near the [001] zone axis of the 2H phase in a gnomonic projection in a plane perpendicular to [001] with viewing angles ±45°. The blue dashed lines indicate the projections of lattice planes (hkl) shown by blue labels. In the chosen projection of the lattice planes, the angles between the Kikuchi bands of planes in the [001] zone directly correspond to angles between the corresponding lattice planes in the crystal structure (i.e. with plane normals perpendicular to [001]). The Kikuchi band edges mark directions of Bragg reflection along the two Kossel cones with their symmetry axis along $g_{hkl}$, and the angle between the band edges is 2 Bragg angles. In the $k_x$ – $k_y$-projection measured in the momentum microscope, the size of the Brillouin zone in the projection plane can be calibrated from the width and orientation of the Kikuchi bands related to $2·|g_{hkl}|$.

In Fig. S6 one can see that the monoclinic "hexagon" (for 300K) is tilted away from the center in the standard setting. It is about 8.5° away from the orthorhombic [001] direction (if the rotation angles are fixed in the simulation), which is due to the different setting of the unit cells.

*Experimental details for transport measurements*

General Electric varnish was used to glue the sample to the Cu sample holder. We use a four-probe geometry with silver electrical contacts (see Fig.**S5**, a). Electric transport measurements were performed in a temperature range of 1.4 to 300 K.

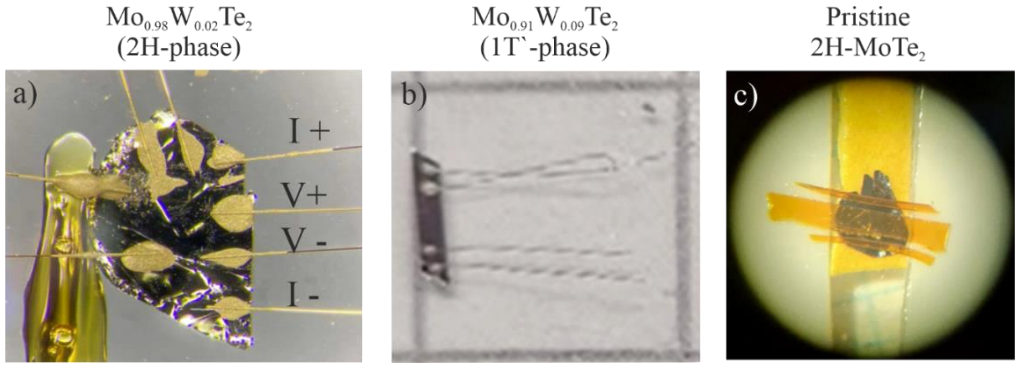

**Figure S7.** a) contacts connection for $Mo_{0.98}W_{0.02}Te_2$ (2H-phase) for transport measurements; b) contacts connection for $Mo_{0.91}W_{0.09}Te_2$ (1T`-phase) along *a* - axis; c) Gold contacts were used for $MoTe_2$.

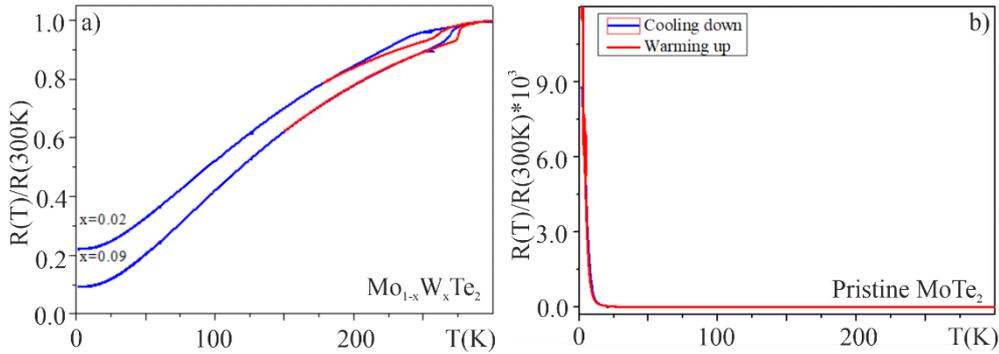

**Figure S8.** Temperature-dependent normalized resistance at zero magnetic field. The hysteresis indicates a first-order 1T` to $T_d$ phase transition.

Bulk $MoTe_2$ exhibits superconductivity with a transition temperature of 0.10 K. Under an external pressure of 11.7 GPa, the transition temperature increases up to a maximum of 8.2 K [7]. Recently, semi-metallic TMDCs have attracted attention due to the discovery of remarkable quantum phenomena. For example, $T_d$ - $WTe_2$ exhibits extremely large magnetoresistance, superconductivity under pressure and a large Nernst effect [76], [77]. Moreover, the 1T`-$MX_2$ monolayer has been predicted to be a two-dimensional topological insulator [2]. The discovery of superconductivity in $T_d$ - $MoTe_2$ (Weyl semimetal phase) may stimulate the exploration of topological superconductivity along with emergent space-time supersymmetry [7], [78].

The EPC calculations in [63] show that the EPC contribution of the electronic self-energy increases with temperature due to the rapid increase in the phonon occupation number and the induced changes in the electronic structure could close the small gap between two bands forming W1 WPs, causing a topological phase transition from one phase (TP II, with 4 WPs) to another (TP I, with 8 WPs). When this transition occurs, the phonon linewidth also reverses. This can explain the anomaly in the linewidths of the $^2A1$ mode.

Published transport measurements reveal that pristine $T_d$ - $MoTe_2$ gradually becomes superconducting below T ~ 0.3 K and zero resistivity is observed at $T_c$ = 0.10 K [7].

Elastoresistance measurements have been performed as described in more details in Ref. [50]. The sample were glued on the top side of a piezoelectric stack (Pst 150/7*5*5 from Piezomechanik Gmbh) and a miniature strain gauge was glued on the back side to monitor the strain variation upon applying voltage to the piezoelectric stack. The crystallographic a-axis of the $Mo_{0.91}W_{0.09}$ sample was aligned with the poling direction of the piezoelectric stack. Given the differential thermal expansion between the piezoelectric and the sample, a small strain is applied to the sample even in the absence of applied voltage to the piezoelectric. However, as can be seen in Fig.S9(a), this effect is small enough that it does not modify the electrical transport property of the sample.

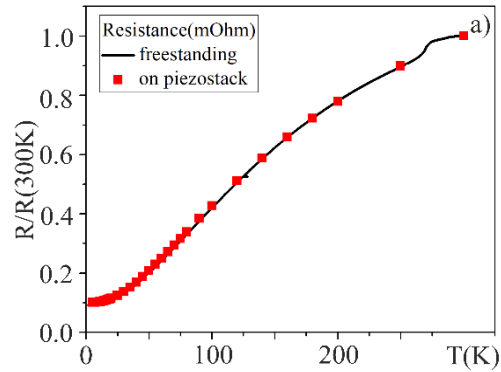

**Figure S9.** Resistance of samples that are freestanding or glued on the piezo-stack sample.